\pdfoutput=1
\documentclass[9pt,a4paper,twocolumn]{article}

\usepackage[utf8]{inputenc}

\usepackage{libertine}
\usepackage[libertine]{newtxmath}

\usepackage{graphicx}
\usepackage[pdfa,pdfproducer={},pdfcreator={}]{hyperref}
\usepackage{url}

\usepackage{mathtools}

\usepackage{xcolor}
\usepackage{colortbl} 

\definecolor{bleudefrance}{rgb}{0.0, 0.28, 0.67}
\definecolor{bleudefrance}{rgb}{0., 0.37, 0.67}

\definecolor{histoblue}{HTML}{3981bf}
\definecolor{histobrown}{HTML}{99886e}
\definecolor{histoyellow}{HTML}{c2c099}

\definecolor{verdevida}{RGB}{108,193,76}
\definecolor{verdevida2}{RGB}{56,100,39}
\hypersetup{
	colorlinks = true,
	citecolor ={bleudefrance},
	allcolors= {bleudefrance},
	linkbordercolor = {white},
}

\usepackage[font=small,justification=RaggedRight]{caption}

\usepackage{pgfgantt}

\usepackage{microtype}
\usepackage[small,compact]{titlesec}

\titleformat{\section}{\bfseries\sffamily\scshape\color{black}}{\arabic{section}}{1em}{\centering\MakeUppercase}

\titleformat{\subsection}{\raggedright\bfseries\sffamily\scshape\small}{\arabic{section}.\arabic{subsection}}{1em}{\MakeUppercase}
\titleformat{\subsubsection}{\centering\bfseries\sffamily\scshape\footnotesize}{\arabic{section}.\arabic{subsection}.\arabic{subsubsection}}{1em}{\MakeUppercase}

\titlespacing{\subsubsection}{0pt}{*4}{*1}
\titlespacing{\subsection}{0pt}{*4}{*1}
\titlespacing{\section}{0pt}{*5}{*2}

\usepackage{fancyheadings}
\usepackage[left=1.50cm, right=1.50cm, top=1.5cm, bottom=2.00cm]{geometry}
\usepackage[backend=bibtex,isbn=true,url=true,eprint=false, bibstyle=numeric,sorting=none,    style=phys,articletitle=false,biblabel=superscript]{biblatex}

\AtEveryBibitem{%
	\clearfield{note}%
}

\DeclareCiteCommand{\citenum}
{}
{\printfield{labelnumber}}
{}
{}

\def\blx@bblfile@bibtex{% instead of ...\blx@bblfile@@biber
	\blx@secinit
	\begingroup
	\blx@bblstart
\begingroup
\makeatletter
\@ifundefined{ver@biblatex.sty}
  {\@latex@error
     {Missing 'biblatex' package}
     {The bibliography requires the 'biblatex' package.}
      \aftergroup }
  {}
\endgroup

\sortlist[entry]{none/global/}
  \entry{skoog2003fundamentals}{inbook}{}
    \name{author}{4}{}{%
      {{hash=SD}{%
         family={Skoog},
         familyi={S\bibinitperiod},
         given={DA},
         giveni={D\bibinitperiod},
      }}%
      {{hash=WD}{%
         family={West},
         familyi={W\bibinitperiod},
         given={DM},
         giveni={D\bibinitperiod},
      }}%
      {{hash=HF}{%
         family={Holler},
         familyi={H\bibinitperiod},
         given={FJ},
         giveni={F\bibinitperiod},
      }}%
      {{hash=CS}{%
         family={Crouch},
         familyi={C\bibinitperiod},
         given={SR},
         giveni={S\bibinitperiod},
      }}%
    }
    \list{publisher}{1}{%
      {Brooks Cole, Stanford}%
    }
    \strng{namehash}{SD+1}
    \strng{fullhash}{SDWDHFCS1}
    \field{labelnamesource}{author}
    \field{labeltitlesource}{title}
  \field{isbn}{\href{https://isbnsearch.org/isbn/9780030355233}{9780030355233}}
    \field{title}{Fundamentals of Analytical Chemistry}
    \field{year}{2003}
  \endentry

  \entry{bare2000}{article}{}
    \name{author}{1}{}{%
      {{hash=BWD}{%
         family={Bare},
         familyi={B\bibinitperiod},
         given={William\bibnamedelima D.},
         giveni={W\bibinitperiod\bibinitdelim D\bibinitperiod},
      }}%
    }
    \strng{namehash}{BWD1}
    \strng{fullhash}{BWD1}
    \field{labelnamesource}{author}
    \field{labeltitlesource}{title}
    \verb{doi}
    \verb 10.1021/ed077p929
    \endverb
    \field{number}{7}
    \field{pages}{929}
    \field{title}{A More Pedagogically Sound Treatment of Beer's Law: A
  Derivation Based on a Corpuscular-Probability Model}
    \field{volume}{77}
    \field{journaltitle}{J. Chem. Ed.}
    \field{year}{2000}
  \endentry

  \entry{Strong}{article}{}
    \name{author}{1}{}{%
      {{hash=SFC}{%
         family={Strong},
         familyi={S\bibinitperiod},
         given={F.\bibnamedelima C.},
         giveni={F\bibinitperiod\bibinitdelim C\bibinitperiod},
      }}%
    }
    \strng{namehash}{SFC1}
    \strng{fullhash}{SFC1}
    \field{labelnamesource}{author}
    \field{labeltitlesource}{title}
    \verb{doi}
    \verb 10.1021/ac60062a020
    \endverb
    \field{number}{2}
    \field{pages}{338\bibrangedash 342}
    \field{title}{Theoretical Basis of Bouguer-Beer Law of Radiation
  Absorption}
    \field{volume}{24}
    \field{journaltitle}{Anal. Chem.}
    \field{year}{1952}
  \endentry

  \entry{Swinehart}{article}{}
    \name{author}{1}{}{%
      {{hash=SDF}{%
         family={Swinehart},
         familyi={S\bibinitperiod},
         given={D.\bibnamedelima F.},
         giveni={D\bibinitperiod\bibinitdelim F\bibinitperiod},
      }}%
    }
    \strng{namehash}{SDF1}
    \strng{fullhash}{SDF1}
    \field{labelnamesource}{author}
    \field{labeltitlesource}{title}
    \verb{doi}
    \verb 10.1021/ed039p333
    \endverb
    \field{number}{7}
    \field{pages}{333}
    \field{title}{The Beer-Lambert Law}
    \field{volume}{39}
    \field{journaltitle}{J. Chem. Ed.}
    \field{year}{1962}
  \endentry

  \entry{Pinkerton1964}{article}{}
    \name{author}{1}{}{%
      {{hash=PRC}{%
         family={Pinkerton},
         familyi={P\bibinitperiod},
         given={Richard\bibnamedelima C.},
         giveni={R\bibinitperiod\bibinitdelim C\bibinitperiod},
      }}%
    }
    \strng{namehash}{PRC1}
    \strng{fullhash}{PRC1}
    \field{labelnamesource}{author}
    \field{labeltitlesource}{title}
    \verb{doi}
    \verb 10.1021/ed041p366
    \endverb
    \field{number}{7}
    \field{pages}{366}
    \field{title}{Beer's law without calculus}
    \field{volume}{41}
    \field{journaltitle}{J. Chem. Ed.}
    \field{year}{1964}
  \endentry

  \entry{Lykos1992}{article}{}
    \name{author}{1}{}{%
      {{hash=LP}{%
         family={Lykos},
         familyi={L\bibinitperiod},
         given={Peter},
         giveni={P\bibinitperiod},
      }}%
    }
    \strng{namehash}{LP1}
    \strng{fullhash}{LP1}
    \field{labelnamesource}{author}
    \field{labeltitlesource}{title}
    \verb{doi}
    \verb 10.1021/ed069p730
    \endverb
    \field{number}{9}
    \field{pages}{730}
    \field{title}{The Beer-Lambert law revisited: A development without
  calculus}
    \field{volume}{69}
    \field{journaltitle}{J. Chem. Ed.}
    \field{year}{1992}
  \endentry

  \entry{Santos}{article}{}
    \name{author}{1}{}{%
      {{hash=BSMN}{%
         family={Berberan-Santos},
         familyi={B\bibinitperiod-S\bibinitperiod},
         given={M.\bibnamedelima N.},
         giveni={M\bibinitperiod\bibinitdelim N\bibinitperiod},
      }}%
    }
    \strng{namehash}{BSMN1}
    \strng{fullhash}{BSMN1}
    \field{labelnamesource}{author}
    \field{labeltitlesource}{title}
    \verb{doi}
    \verb 10.1021/ed067p757
    \endverb
    \field{number}{9}
    \field{pages}{757}
    \field{title}{Beer's law revisited}
    \field{volume}{67}
    \field{journaltitle}{J. Chem. Ed.}
    \field{year}{1990}
  \endentry

  \entry{Daniels}{article}{}
    \name{author}{1}{}{%
      {{hash=DRS}{%
         family={Daniels},
         familyi={D\bibinitperiod},
         given={R.\bibnamedelima Scott},
         giveni={R\bibinitperiod\bibinitdelim S\bibinitperiod},
      }}%
    }
    \strng{namehash}{DRS1}
    \strng{fullhash}{DRS1}
    \field{labelnamesource}{author}
    \field{labeltitlesource}{title}
    \verb{doi}
    \verb 10.1021/ed076p138
    \endverb
    \field{number}{1}
    \field{pages}{138}
    \field{title}{A Random Number Model for Beer's Law - Atom Shadowing}
    \field{volume}{76}
    \field{journaltitle}{J. Chem. Ed.}
    \field{year}{1999}
  \endentry

  \entry{klein2016handbook}{book}{}
    \name{author}{4}{}{%
      {{hash=KJP}{%
         family={Klein},
         familyi={K\bibinitperiod},
         given={John\bibnamedelima P},
         giveni={J\bibinitperiod\bibinitdelim P\bibinitperiod},
      }}%
      {{hash=VHHC}{%
         family={Van\bibnamedelima Houwelingen},
         familyi={V\bibinitperiod\bibinitdelim H\bibinitperiod},
         given={Hans\bibnamedelima C},
         giveni={H\bibinitperiod\bibinitdelim C\bibinitperiod},
      }}%
      {{hash=IJG}{%
         family={Ibrahim},
         familyi={I\bibinitperiod},
         given={Joseph\bibnamedelima G},
         giveni={J\bibinitperiod\bibinitdelim G\bibinitperiod},
      }}%
      {{hash=STH}{%
         family={Scheike},
         familyi={S\bibinitperiod},
         given={Thomas\bibnamedelima H},
         giveni={T\bibinitperiod\bibinitdelim H\bibinitperiod},
      }}%
    }
    \list{publisher}{1}{%
      {CRC Press}%
    }
    \strng{namehash}{KJP+1}
    \strng{fullhash}{KJPVHHCIJGSTH1}
    \field{labelnamesource}{author}
    \field{labeltitlesource}{title}
  \field{isbn}{\href{https://isbnsearch.org/isbn/9781466555679}{9781466555679}}
    \field{title}{Handbook of survival analysis}
    \field{year}{2016}
  \endentry

  \entry{last2017lectures}{book}{}
    \name{author}{2}{}{%
      {{hash=LG}{%
         family={Last},
         familyi={L\bibinitperiod},
         given={G{\"u}nter},
         giveni={G\bibinitperiod},
      }}%
      {{hash=PM}{%
         family={Penrose},
         familyi={P\bibinitperiod},
         given={Mathew},
         giveni={M\bibinitperiod},
      }}%
    }
    \list{publisher}{1}{%
      {Cambridge University Press}%
    }
    \strng{namehash}{LGPM1}
    \strng{fullhash}{LGPM1}
    \field{labelnamesource}{author}
    \field{labeltitlesource}{title}
  \field{isbn}{\href{https://isbnsearch.org/isbn/9781107458437}{9781107458437}}
    \field{title}{Lectures on the Poisson process}
    \field{volume}{7}
    \field{year}{2017}
  \endentry

  \entry{Lohman}{article}{}
    \name{author}{1}{}{%
      {{hash=LFH}{%
         family={Lohman},
         familyi={L\bibinitperiod},
         given={Fred\bibnamedelima H.},
         giveni={F\bibinitperiod\bibinitdelim H\bibinitperiod},
      }}%
    }
    \strng{namehash}{LFH1}
    \strng{fullhash}{LFH1}
    \field{labelnamesource}{author}
    \field{labeltitlesource}{title}
    \verb{doi}
    \verb 10.1021/ed032p155
    \endverb
    \field{number}{3}
    \field{pages}{155}
    \field{title}{The mathematical combination of Lambert's law and Beer's law}
    \field{volume}{32}
    \field{journaltitle}{J. Chem. Ed.}
    \field{year}{1955}
  \endentry

  \entry{Bird}{inbook}{}
    \name{author}{4}{}{%
      {{hash=BRB}{%
         family={Bird},
         familyi={B\bibinitperiod},
         given={R.\bibnamedelima Byron},
         giveni={R\bibinitperiod\bibinitdelim B\bibinitperiod},
      }}%
      {{hash=SWE}{%
         family={Stewart},
         familyi={S\bibinitperiod},
         given={Warren\bibnamedelima E.},
         giveni={W\bibinitperiod\bibinitdelim E\bibinitperiod},
      }}%
      {{hash=LEN}{%
         family={Lightfoot},
         familyi={L\bibinitperiod},
         given={Edwin\bibnamedelima N.},
         giveni={E\bibinitperiod\bibinitdelim N\bibinitperiod},
      }}%
      {{hash=KDJ}{%
         family={Klingenberg},
         familyi={K\bibinitperiod},
         given={Daniel\bibnamedelima J.},
         giveni={D\bibinitperiod\bibinitdelim J\bibinitperiod},
      }}%
    }
    \list{publisher}{1}{%
      {Wiley New York}%
    }
    \strng{namehash}{BRB+1}
    \strng{fullhash}{BRBSWELENKDJ1}
    \field{labelnamesource}{author}
    \field{labeltitlesource}{title}
  \field{isbn}{\href{https://isbnsearch.org/isbn/978111895372-3}{9781118953723}}
    \field{title}{Introductory Transport Phenomena}
    \field{year}{2015}
  \endentry

  \entry{goldbook}{book}{}
    \true{moreauthor}
    \name{author}{2}{}{%
      {{hash=MAD}{%
         family={McNaught},
         familyi={M\bibinitperiod},
         given={Alan\bibnamedelima D},
         giveni={A\bibinitperiod\bibinitdelim D\bibinitperiod},
      }}%
      {{hash=WA}{%
         family={Wilkinson},
         familyi={W\bibinitperiod},
         given={Andrew},
         giveni={A\bibinitperiod},
      }}%
    }
    \list{publisher}{1}{%
      {Blackwell Science Oxford}%
    }
    \strng{namehash}{MADWA+1}
    \strng{fullhash}{MADWA+1}
    \field{labelnamesource}{author}
    \field{labeltitlesource}{title}
    \field{title}{{IUPAC. C}ompendium of chemical terminology.
  (\href{https://doi.org/10.1351/goldbook}{Online version (2019-)})}
    \field{year}{1997}
  \endentry

  \entry{grossman1972non}{book}{}
    \name{author}{2}{}{%
      {{hash=GM}{%
         family={Grossman},
         familyi={G\bibinitperiod},
         given={Michael},
         giveni={M\bibinitperiod},
      }}%
      {{hash=KR}{%
         family={Katz},
         familyi={K\bibinitperiod},
         given={Robert},
         giveni={R\bibinitperiod},
      }}%
    }
    \list{publisher}{1}{%
      {Lee Press}%
    }
    \strng{namehash}{GMKR1}
    \strng{fullhash}{GMKR1}
    \field{labelnamesource}{author}
    \field{labeltitlesource}{title}
  \field{isbn}{\href{https://isbnsearch.org/isbn/9780912938011}{9780912938011}}
    \field{title}{Non-Newtonian Calculus: A Self-contained, Elementary
  Exposition of the Authors' Investigations.}
    \field{year}{1972}
  \endentry

  \entry{GILLESPIE1992}{article}{}
    \name{author}{1}{}{%
      {{hash=GDT}{%
         family={Gillespie},
         familyi={G\bibinitperiod},
         given={Daniel\bibnamedelima T.},
         giveni={D\bibinitperiod\bibinitdelim T\bibinitperiod},
      }}%
    }
    \strng{namehash}{GDT1}
    \strng{fullhash}{GDT1}
    \field{labelnamesource}{author}
    \field{labeltitlesource}{title}
    \verb{doi}
    \verb https://doi.org/10.1016/0378-4371(92)90283-V
    \endverb
    \field{number}{1}
    \field{pages}{404 \bibrangedash  425}
    \field{title}{A rigorous derivation of the chemical master equation}
    \field{volume}{188}
    \field{journaltitle}{Physica A}
    \field{year}{1992}
  \endentry
\endsortlist

	\blx@bblend
	\endgroup
	\csnumgdef{blx@labelnumber@\the\c@refsection}{0}%
	\iftoggle{blx@reencode}{\blx@reencode}{}}

\usepackage{ragged2e}
\usepackage{booktabs}
\usepackage{tabularx}

\usepackage{placeins}

\usepackage[misc]{ifsym}

\usepackage{float}
\restylefloat{table}

\usepackage{appendix} 

\usepackage{abstract}
\usepackage{orcidlink}

\begin{document}

	\title{\Large\textsf{\textbf{\MakeUppercase{{Seven derivations of the Lambert-Beer law}}}}}
	\author{\small \textsf{\scshape{Hernán R. Sánchez}}  \orcidlink{0000-0003-1058-3396} \\   }	
	\date{}
	
	\twocolumn[
	\maketitle
	\begin{center}
	{\vspace{-0.5cm}\small Centro de Química Inorgánica. CONICET  La Plata - UNLP, Argentina.	
		
		\footnotesize {\scriptsize\Letter} \textsf{\href{mailto:hernan.sanchez@quimica.unlp.edu.ar}{hernan.sanchez@quimica.unlp.edu.ar}}}
\end{center}
	\begin{onecolabstract}
 \vspace{-0.3cm}
       Seven derivations of the Lambert-Beer law are proposed in this paper. They were designed to be simple and intuitive. Most of them are suitable for  the classroom. Readers can also benefit from them by looking at the phenomenon from different perspectives, which gives valuable resources when explaining it in class.
  		\vspace*{0.8cm}
	\end{onecolabstract}	
	]

\section{Introduction}
Electromagnetic radiation beams are attenuated by passing through an absorbent material. The Lambert-Beer (L-B) law describes how this attenuation depends on the concentration of the absorbent particles and on the optical path, provided that certain conditions are met\supercite{skoog2003fundamentals}. This work has two principal aims: firstly to provide simple yet rigorous derivations of the L-B law useful to be taught in the classroom. Secondly, to broaden our current understanding of this law by approaching it from different viewpoints.

Many derivations of the L-B law have been proposed\supercite{bare2000,Strong,Swinehart,Pinkerton1964,Lykos1992,Santos,Daniels}. From an abstract point of view, the derivations at some point state a relationship between internal transmittance ($T$) and concentration ($c$) or optical path ($b$) satisfied only by the exponential function. For this, different approaches  can be followed. In \S \ref{Sec: Deduccion 1} a very brief and simple derivation of the L-B law is proposed. What makes this derivation accessible is that the relationship employed is the exponential identity $a^{x+y}=a^{x}a^{y}$ which is simple and known from introductory courses.

Berberan-Santos\supercite{Santos} and Daniels\supercite{Daniels} proposed proofs closely connected to gas kinetic theory. These are pretty rigorous and  provided a clear picture of the phenomenon. A derivation on the same lines is proposed in \S \ref{Sec: Deduccion 2}. It differs from the ones just mentioned in that it is mathematically much simpler. 

In the above derivations a photon is absorbed whenever it finds an absorbing particle.  This is not true under the alternative picture employed in the derivation proposed in \S \ref{Sec: Deduccion 3}.

The above derivations have a strong probabilistic approach involving spatial variables. The latter are related to the width of solution layers or the position of absorbing particles. The derivations proposed in \S \S \ref{Sec: Deduccion 4}  and \ref{Sec: Deduccion 5} maintain the probabilistic approach but  move the focus to the time that a photon remains unabsorbed.  That of  \S \ref{Sec: Deduccion 4}  uses basic elements of survival analysis\supercite{klein2016handbook} and chemical kinetics, while the one from  \S \ref{Sec: Deduccion 5} models the phenomenon as a Poisson process\supercite{last2017lectures}.

The derivations based in calculus often take the form
\begin{equation*}
\frac{\partial T(x,y)}{\partial x}= k_y y T(x,y)
\end{equation*}
\noindent where $(x,y)$ represents $(b,c)$ or $(c,b)$ and $k_y$ is a proportionality constant. Commonly, radiant power ($P$) or light intensity are used in place of internal transmittance but the relation above can be obtained by a simple change of variables. Both alternatives, that is $(x,y)=(b,c)$ and $(x,y)=(c,b)$, can be combined in one system of two equations which also leads to the L-B law\supercite{Lohman}.

According to Bare,  the derivations found in most undergraduate texts require the use of calculus concepts, and  these  start by considering that the absorption in a layer of infinitesimal width is proportional to that width, for example in the following form: $dP= k_c c P db$\supercite{bare2000}.  A rigorous derivation requires staring from self-evident premises. He argued that it is not obvious what absorption properties an infinitesimally thin film should have because we have no physical experience with such a film.  Bare stressed that the linear relationship is valid only for infinitesimal widths\supercite{bare2000}, and implicitly criticized the approach of Lykos who assumed\supercite{Lykos1992} this for thin but finite layers. Berberan-Santos implied that it is not clear what the width of those layers is and figured it to be similar to that of a molecule\supercite{Santos}. What is clear is that it is not obvious to students how to interpret something on which experts disagree. In \S \ref{Sec: Deduccion 6} a calculus-based derivation is provided. It reveals and highlight the probabilistic nature of the process and  does not begin by making assumptions about the properties of thin layers.

Finally,  a derivation based on the continuity equation\supercite{Bird} is given in \S \ref{Sec: Deduccion 7}. It could be considered as a straightforward calculus-based derivation  free from the issues mentioned above.

\section{Proposed derivations}
The transmittance is defined as the ratio of the transmitted radiant power to that incident on the sample\supercite{goldbook}. Among of the many mechanisms of energy loss only that of absorption is considered here. Specifically, absorption due to processes involving a single photon, which are the most common for reasonably low radiant powers. We will consider that the sample is homogeneous and isotropic and that the incident radiation is monochromatic, collimated and normal to the surface of the sample. In such case, as  electromagnetic radiation can be thought as a stream of photons, transmittance equals the fraction of non-absorbed photons. This corresponds to the probability that a randomly chosen photon will pass through the sample. Thus, this probability equals the transmittance and the problem may be stated as finding how the former depends on the path length and the concentration  of the absorbing particles, lets call this function $\mathcal{P}(c,b)$. These considerations will be made in the most of the proposed derivations.
\subsection{First derivation}\label{Sec: Deduccion 1}
The process of a photon traveling through the sample of length $b_1 + b_2$ can be subdivided into two processes, one corresponding to the photon traveling in the first section (of length $b_1$) and the other to the photon traveling in the remaining section (of length $b_2$). For each section, there are two possibilities: the photon passes through or does not. 

Let $\mathcal{P}(c, b_1)$ be the probability that the photon passes through the first section. The probability of the photon passing through the second is dependent on the latter: if the first one is not crossed the photon could never cross the second one. In fact, this probability equals the probability that the photon passes through the whole sample. Lets denote by $\mathcal{P}(c,b_2|c,b_x>b_1)$ to the probability that the photon has crossed the second section since we know that the first section was crossed.  The probability that the photon passes through the sample, $\mathcal{P}(c,b_1+b_2)$, will then be  $\mathcal{P}(c,b_1+b_2)=\mathcal{P}(c,b_1)\mathcal{P}(c,b_2|c,b_x>b_1)$ according to the chain rule of probability. The fact that a photon has passed through a section does not change its propensity to be absorbed in the future. This implies that $\mathcal{P}(c,b_2|c,b_x>b_1)$ must be equal to the probability of a photon passing a first section of length $b_2$, i.e. $\mathcal{P}(c,b_2|c,b_x>b_1)=\mathcal{P}(c,b_2)$, then
\begin{equation}\label{Eq: P(b1+b2)=P(b1)P(b2)}
\mathcal{P}(c,b_1+b_2) = \mathcal{P}(c,b_1)\mathcal{P}(c,b_2)
\end{equation}

The only non-trivial continuous real function that satisfies the above equality for a fixed $c$ is the exponential function\footnote{$\mathcal{P}(c,b)=1$ and $\mathcal{P}(c,b)=0$ are trivial and nonphysical solutions.} $\mathcal{P}(c,b)=a^{f_1(c) b}$. In fact, this is one of the many ways in which the exponential function can be characterized.

We will analyze the dependence on concentration. To model the process we will consider the following: the photon is absorbed when it collides with an absorbing particle that remains unmodified during that process. Thus, in order for the photon not to be absorbed, no absorbing particle can be in the region of space ($\mathcal{V}$) where the photon would be obstructed. Then, the probability that it will not be absorbed is equal to the probability that there are no absorbent particles in that region. If they are divided into two subsets built randomly, the sample concentration  $c$ may be expressed as the sum of the two concentrations ($c_1$ and $c_2$) due to  both subsets. Let $\mathcal{P}(c_1+c_2,b)$ be the probability that there are no particles in $\mathcal{V}$. No particles in $\mathcal{V}$ implies that there are no particles contributing to $c_1$ and no particles contributing to $c_2$ in that region. The probabilities of these events are $\mathcal{P}(c_1,b)$ and $\mathcal{P}(c_2,b)$, respectively. A good approximation is to consider that the absorbing particles are independent of each other and follow a uniform distribution imposed by the homogeneity constraint. Thus, the two events are independent and the probability of them both occurring equals the product of their probabilities
\begin{equation}\label{Eq: P(c1+c2)=P(c1)P(c2)}
\mathcal{P}(c_1+c_2,b) = \mathcal{P}(c_1,b)\mathcal{P}(c_2,b)
\end{equation}

\noindent which implies $\mathcal{P}(c,b)=a^{f_2(b)c}$. Then, as $\mathcal{P}(c,b)=a^{f_1(c)b}=a^{f_2(b)c}$ it follows that 
\begin{equation*}
\mathcal{P}(c,b) = T(c,b)=a^{\beta b c}
\end{equation*}
\noindent which is the L-B law and $\beta$ is constant. Because probabilities always lie between 0 and 1, $0\leq  a^{\beta}\leq 1$.  In real systems $0< a^{\beta}< 1$.

\subsection{Second derivation}\label{Sec: Deduccion 2}
Following the logic used to obtain Eq. \ref{Eq: P(c1+c2)=P(c1)P(c2)}, we can build as many subsets as absorbent particles. If $c_0$  represents the concentration corresponding to having only one particle in the sample, $N$ represents the number of particles and $A$ the cross section of the sample
\begin{equation*}
\mathcal{P}(c,b)=\prod _i ^N \mathcal{P}(c_0,b) = \mathcal{P}(c_0,b)^N = \mathcal{P}(c_0,b)^{Abc}
\end{equation*}
$\mathcal{P}(c_0 , b)$ does not depend on $b$ provided that $A$ remains the same, so the proof was completed.

\subsection{Third derivation}\label{Sec: Deduccion 3}
 For the present derivation an alternative picture will be used. Again, it will be considered that the absorbing particles must be in $\mathcal{V}$ for the photon to be absorbed. In contrast to the previous picture, a particle in  $\mathcal{V}$ does not implies that the absorption will necessarily occur, but there will be a probability $\mathcal{P}_{2}$ of it occurring.

For a given absorbing particle, the probability ($\mathcal{P}_{1}$) of being in $\mathcal{V}$ is the the fraction of the total volume that  $\mathcal{V}$ represents. Having $N$ particles, the probability of having $k$ of them in $\mathcal{V}$ follows the Bernoulli distribution
\begin{equation*}
\mathcal{P}(k,N;\mathcal{P}_{1})=\binom{N}{k}\mathcal{P}_{1}^k(1-\mathcal{P}_{1})^{N-k}
\end{equation*}
With $k$ particles in $\mathcal{V}$, the probability of $m$ of them having the capacity to absorb the photon in case the encounter takes place is also given by the Bernoulli distribution
\begin{equation*}
\mathcal{P}(m,k;\mathcal{P}_{2})=\binom{k}{m}\mathcal{P}_{2}^m(1-\mathcal{P}_{2})^{k-m}
\end{equation*}
Thus, the probability of having $m$ particles capable of absorbing the photon is
\begin{equation*}
\mathcal{P}(m,N;\mathcal{P}_{1},\mathcal{P}_{2})=\sum_{k=m}^{N}\binom{N}{k}\binom{k}{m}\mathcal{P}_{1}^k(1-\mathcal{P}_{1})^{N-k}\mathcal{P}_{2}^m(1-\mathcal{P}_{2})^{k-m}
\end{equation*}
Through algebraic transformations it can be simplified to 
\begin{equation*}
\mathcal{P}(m,N;\mathcal{P}_{1}\times\mathcal{P}_{2})=\binom{N}{m}(\mathcal{P}_{1}\mathcal{P}_{1})^m(1-\mathcal{P}_{1}\mathcal{P}_{2})^{N-m}
\end{equation*}
The probability of the photon passing trough the sample equals the above expression for $m=0$, 
\begin{equation*}
\mathcal{P}(c,b)=\mathcal{P}(0,N;\mathcal{P}_{1}\times\mathcal{P}_{2})=(1-\mathcal{P}_{1}\mathcal{P}_{2})^{N}=(1-\mathcal{P}_{1}\mathcal{P}_{2})^{Abc}
\end{equation*}

By replacing probability with transmittance this expression turns into the Lambert-Beer law. 

\subsubsection{Mathematical variation}
This treatment can be simplified by analyzing the particles one by one. For the photon not to be absorbed, each absorbing particle must satisfy one of the two following conditions: be outside of $\mathcal{V}$ (whose probability is $1-\mathcal{P}_{1}$), or be inside of $\mathcal{V}$  and not to absorb (of which the probability is $\mathcal{P}_{1}\times(1-\mathcal{P}_{2})$ as these are independent events). Because these are mutually exclusive the probability of one of them occurring is
\begin{equation*}
(1-\mathcal{P}_{1}) + \mathcal{P}_{1}\times(1-\mathcal{P}_{2}) = 1-\mathcal{P}_{1}\mathcal{P}_{2}
\end{equation*} 
The occurrence of the above events does not depend on the remaining particles, then for the whole sample
\begin{equation*}
(1-\mathcal{P}_{1}\mathcal{P}_{2})^N
\end{equation*} 
The rest of the proof is exactly the same as the previous one.

\subsection{Fourth derivation}\label{Sec: Deduccion 4}
The problem will be approached from a temporal rather than a spatial perspective. Consider the event of a photon being absorbed. Let $\tau$ be a non-negative random variable denoting the waiting time until this (eventually) occurs, and $f(t)$ the corresponding probability density function. The survival function, $S(t)$, is the probability of the photon remaining unabsorbed until a time $t$. If we set $t=0$ when the photon enters the sample
\begin{equation*}
S(t)= 1 - \int _0 ^t f(t^\dagger) dt^\dagger
\end{equation*}
This implies that $f(t)=-S^\prime(t)$. The distribution of $\tau$ can also be characterized though the hazard function, $h(t)$, defined by
\begin{equation}\label{eq: definition hazard}
h(t)=\lim_{\Delta t\to 0} \frac{\mathcal{P}(t \le \tau < t+\Delta t\mid \tau\ge t)}{\Delta t}
\end{equation}
The latter represents the instantaneous rate of occurrence of the event. The absorption of photons can be treated as an elemental chemical reaction between the photons and the absorbent species, being of first-order in each of them provided that light intensity is reasonably low.  Then,  as long as the photon is inside of the solution, $h(t)$ is proportional to the reaction rate with which the absorption can be modeled. And the concentration of photons is constant because it is due to a single photon. Then, $h(t)=\kappa c$ where $\kappa$ is a proportionality constant. Notice that the numerator in Eq. \ref{eq: definition hazard} can be rewritten as $f(t)dt/S(t)$. Thus, due to $f(t)=-S^\prime(t)$ we have
\begin{equation*}
h(t)= -\frac{S^\prime(t)}{S(t)}= - \frac{d\ln S(t)}{dt}
\end{equation*}
\noindent then
\begin{equation*}
S(t)= e^{-\int_0 ^t h(t^\dagger)dt^\dagger}=e^{-\int_0 ^t \kappa c dt^\dagger}=e^{-\kappa ct}
\end{equation*}
The time $t$ is related to the distance traveled by a photon, $b$,  and its speed, $v_x$, through $t=b/v_x$. This brings out the equality $S(t)=T(c,b)$ for a large set of photons, then
\begin{equation*}
T(b,c)= e^{-\frac{\kappa}{v_x} b c}
\end{equation*}
\noindent which is the L-B law.

\subsection{Fifth derivation}\label{Sec: Deduccion 5}
Consider a single photon. Let us imagine that once its absorption has taken place the photon is not altered and can be absorbed again indefinitely. This is a false assumption, but it will help us to model the phenomenon as a Poisson process. Using the former is justified because we will only consider what happened in times prior to the first eventual absorption.  As a photon travels through the solution its chance of being absorbed is the same for any point in time. If once again the process is modeled as a chemical reaction, the number of occurrences per unit time ($\lambda$) would be proportional to the reaction rate. It can be written as $\lambda=\kappa c$. The expected number of hypothetical absorption during a time $t$  equals $\lambda t$. Two other remarkable features are that the hypothetical absorptions are independent of each other, and that only one absorption can occur at a given time. This is enough to model the phenomenon as a Poisson process.

The number of hypothetical absorptions can be represented with a random variable ($X$) that follows the Poisson distribution. The latter is a discrete probability distribution that can be used for modeling the number of times ($k$) an event (in this case absorption) takes place in a fixed time interval.  The corresponding probability mass function is
\begin{equation}
\nonumber \mathcal{P}(X=k) = \left\{
\begin{array}{l l}
\frac{ (\lambda t)^k e^{-\lambda t}}{k!}& \quad \text{for  } k \in \{0,1,2,\dots\}\\
0  & \quad \text{ otherwise}
\end{array} \right.
\end{equation}
The probability that no absorption takes place ($k=0$) up to a time $t$ is 
\begin{equation}
\mathcal{P}(X=0)=\frac{(\lambda t) ^0 e^{-\lambda t}}{0!}=e^{-\kappa  ct}
\end{equation}
The proof concludes in the same way as that proposed in \S \ref{Sec: Deduccion 4}.

\subsection{Sixth derivation}\label{Sec: Deduccion 6}
Consider a set of $n$ layers at positions $b_1,b_2,\dots b_n$ where $b_j>b_i$ if $j>i$ and let $b_0=0$. From Eq. \ref{Eq: P(b1+b2)=P(b1)P(b2)} we have that $\mathcal{P}(c,b_k)= \mathcal{P}(c,b_{k-1})\mathcal{P}(c,b_k-b_{k-1})$. It can be rewritten as
\begin{equation*}
\mathcal{P}(c,b_k-b_{k-1}) = 1 - \frac{\Delta \mathcal{P}_k}{\mathcal{P}(c,b_{k-1})}
\end{equation*}
\noindent where $\Delta \mathcal{P}_k:=\mathcal{P}(c,b_{k-1})-\mathcal{P}(c,b_k)>0$. From this and Eq. \ref{Eq: P(b1+b2)=P(b1)P(b2)} the probability of a photon passing through the whole sample is the product of the probability of passing through each layer of width $b_k-b_{k-1}$,
\begin{equation*}
\prod _{k=1}^n \mathcal{P}(c,b_k-b_{k-1})=\prod _{k=1}^n \left(1 - \frac{\Delta \mathcal{P}_k}{\mathcal{P}(c,b_{k-1})}\right)
\end{equation*}
As $e^x$ can be expanded in Maclaurin series as $1+x+\mathcal{O}(x^2)$, for small $\Delta \mathcal{P}_k$ we have
\begin{equation*}
1 - \frac{\Delta \mathcal{P}_k}{\mathcal{P}(c,b_{k-1})} \sim  e^{ -\frac{\Delta \mathcal{P}_k}{\mathcal{P}(c,b_{k-1})}}
\end{equation*}
\noindent then
\begin{align*}
\mathcal{P}(c,b)&=\lim _{n\to \infty} \prod _{k=1}^n e^{ -\frac{\Delta \mathcal{P}_k}{\mathcal{P}(c,b_{k-1})}}=\lim _{n\to \infty}e^{\ln [\prod _{k=1}^n e^{- \frac{\Delta \mathcal{P}_k}{\mathcal{P}(c,b_{k-1})}} ] }\\
&=\lim _{n\to \infty} e^{-\sum _{k=1} ^n \left( \ln   e^{ \frac{1}{\mathcal{P}(c,b_{k-1})}}  \Delta \mathcal{P}_k\right) }= e^{-\int _1 ^{\mathcal{P}} \ln  \left(e^{ \frac{1}{\mathcal{P}^\dagger}}\right)   d\mathcal{P}^\dagger }\\
&= e^{-\int _1 ^{\mathcal{P}} \frac{1}{\mathcal{P}^\dagger}    d\mathcal{P}^\dagger }
\end{align*}
\noindent that can be rewritten in terms of $P$ and $P_0:=P(c,0)$ as
\begin{equation}
e^{-\int _1 ^{\mathcal{P}} \frac{1}{\mathcal{P}}    d\mathcal{P} }= e^{-\int _{P_0} ^{P} \frac{1}{P^\dagger}    dP^\dagger }
\end{equation}
A change corresponding to a proportion $\mathcal{P}(c,b_k-b_{k-1})$ can be attributed to the existence of  ~$~\Delta~N~=~A~c~(~b_k~-~b_{k-1}~)$ absorbing particles. Thus, there must be a function of $\Delta N$, $f(\Delta N)$, that equals $1 -\Delta \mathcal{P}_k/\mathcal{P}(c,b_{k-1})$. We know about $f(\Delta N)$ that $\lim_{\Delta N\to 0^+} f(\Delta N)=1$. Expanding in Taylor series around zero and for small $\Delta N$, 
\begin{equation*}
\lim _{\Delta N\to 0^+} f(\Delta N) = \lim _{\Delta N\to 0^+} 1 + \frac{df(\Delta N)}{d\Delta N} \bigg\rvert_{\Delta N=0} \Delta N =\lim _{\Delta N\to 0^+}  \left(e^{\alpha}\right)^{\Delta N}
\end{equation*}
\noindent where $\alpha:=f^\prime(0)$.
For the whole sample, performing steps analogous to the previous ones 
\begin{equation*}
\lim _{n\to \infty} \prod_{k=1}^n f(\Delta N)=\lim _{n\to \infty}  \prod_{k=1}^n  \left(e^{\alpha}\right)^{\Delta N}=e^{\int_{0}^{N}\alpha dN   } = e^{\int_{0}^{b}\alpha A c db^\dagger   } 
\end{equation*}
Thus, 
\begin{equation*}
e^{-\int _{P_0} ^{P} \frac{1}{P}    dP }=e^{\int_{0}^{b}\alpha A c db^\dagger   }
\end{equation*}
\noindent Taking logarithms of both sides the staring point of the standard calculus-based derivations is obtained, from where the L-B law can be derived.

\subsection{Seventh derivation}\label{Sec: Deduccion 7}
Let $\phi$ be the volume density of photons in the sample and let $\mathbf{v}=[v_x,v_y,v_z]^T$ be their velocity field. Notice that  $\phi$ depends on the position and, eventually, of time. The flux of photons in the sample is defined by $\mathbf{J}=[J_x,J_y,J_z]^T:=\phi\mathbf{v}$. Let $Q$ represents the number of photons absorbed per unit volume per unit time. According to the continuity equation in its differential form 
\begin{equation*}
\frac{\partial \phi}{\partial t} + \nabla \cdot \mathbf{J} + Q=0
\end{equation*}
If  we again model  the absorption process as an elemental chemical reaction: $Q=k c\phi$ where $k$ is the reaction rate constant.  Being that the incident radiation is collimated we will consider that light moves along the $x$-axis of a cartesian coordinate system, which implies that $\nabla \cdot \mathbf{J}=\frac{\partial J_x}{\partial x}$. In steady state $\partial \phi/\partial t =0$, then
\begin{equation*}
\frac{\partial J_x}{\partial x}=-k c \phi=- \frac{k  c (v_x\phi)}{v_x} =-\frac{k  c J_x}{v_x}
\end{equation*}
If $J_x$ is multiplied by the product between the frequency of the radiation ($\nu$) and the Planck's constant ($h$), the $x-$component of the energy flux is obtained. Thus, multiplying by $h\nu$ and considering that the derivative is a linear map

\begin{equation*}
h\nu\frac{\partial J_x }{\partial x}= \frac{\partial (h\nu J_x )}{\partial x} =-\frac{k  c J_x h\nu}{v_x}\\
\end{equation*}

The radiant power can be obtained by integrating the energy flux with respect to the transversal area. To write the above equation in terms of the radiant power we do
\begin{equation*}
\iint_S\frac{\partial (J_x h\nu)}{\partial x}dydz = \iint_S -\frac{k  c }{v_x}  J_x h\nu dydz
\end{equation*}
and  then we employ the Leibniz rule
\begin{align*}
\frac{ d (\iint_S J_x h\nu dydz)}{d x} &= -\frac{k  c }{v_x} \iint_S J_x h\nu dydz\\
\frac{ d P}{d x} &= -\frac{k  c }{v_x} P
\end{align*}
\noindent from where the usual steps of the standard derivation can be followed.

\section{Discussion}
Many proofs were proposed in the previous section. Their relative advantages depend on the course in question. The proof proposed in \S \ref{Sec: Deduccion 2} is difficult to beat in terms of brevity. Originally this proof was to be discussed in an extended form. However, in a subsequent literature search conducted for this paper, I noticed that it shares many similarities with the works of Berberan-Santos\supercite{Santos} and Daniels\supercite{Daniels}. Nevertheless, the proof was described here because it could be useful for some since it is much simpler mathematically. Berberan-Santos implicitly treated the process of including or not including a given particle in $\mathcal{V}$ as a Bernoulli trial and used the Binomial Law and the well-known equation
\[ e^x = \lim_{n\to\infty}\left(1 + \frac{x}{n}\right)^{n}\]
His proof has been considered complex\supercite{Lykos1992,bare2000}. Daniels' solution makes use of geometric series which for many students is not trivial. The simplicity of the proposed derivation can make it more accessible while maintaining the same rigor and picture which is nicely described in the early works.

The proof proposed in \S \ref{Sec: Deduccion 1} is almost as simple and brief as that of \S \ref{Sec: Deduccion 2}. It makes use of conditional probability when dealing with the dependence on the optical path. An alternative approach is to consider that the sample is split into independent layers and that for each of them there is a photon (all of them with the same frequency) trying to pass through with some probability. The probability of a photon (with that frequency) passing through the whole sample equals the product of those probabilities. This seems to be the idea behind the approach of Bare\supercite{bare2000}. Making this clear, his derivation becomes another simple alternative. 

In the derivations proposed above (except those of \S\S \ref{Sec: Deduccion 5}, \ref{Sec: Deduccion 6}  and \ref{Sec: Deduccion 7}), and in those from the early works, it was considered that a photon is absorbed whenever it encounters an absorbing particle. The absorption properties of each particle are attributed to a wavelength-dependent cross-section not directly related to the size of the particle. This is a valid approach, though perhaps not the easiest to imagine. For example, it may be counterintuitive when comparing solute and solvent in a colored diluted aqueous solution under visible light. The proofs found in \S \ref{Sec: Deduccion  3} employ an alternative and probably more intuitive picture, so they can be considered valuable alternatives. They generalize the proof given in \S \ref{Sec: Deduccion 2}, the latter corresponds to the particular case where $\mathcal{P}_2=1$. It can be further generalized with relative simplicity to consider more than one type of absorbent particles. This can be done by using the Poisson binomial distribution, but this derivation will be omitted to conserve space.

The derivation proposed in \S \ref{Sec: Deduccion 4}  makes use of fundamental concepts of survival analysis and chemical kinetics. The hazard function was used solely to include the dependence on concentration. For obtaining only the dependence on the optical path,  we could consider the following. In \S \ref{Sec: Deduccion 1} it was mentioned that the fact that a photon has passed through a given layer does not change its propensity to be absorbed in the future. This suggest a memoryless process, that is $\mathcal{P}(\tau>t_1)=\mathcal{P}(\tau>t_1+t_0|\tau>t_0)$. Using the chain rule $\mathcal{P}(\tau>t_1)= \mathcal{P}(\tau>t_1+t_0)/\mathcal{P}(\tau>t_0)$. Then, due to the definition of the survival function: $S(t_0)S(t_1)=S(t_1+t_0)$ which implies that $S$ is an exponential function of time, therefore, of the optical path.

The proof provided in  \S \ref{Sec: Deduccion 5}  may seem challenging, but the idea behind it is pretty simple. For a given photon, the absorption probability density is constant in time. This implies that, if the photon could be adsorbed indefinitely, the variable representing the times at which the absorptions occur follows a uniform distribution. The count of (hypothetical) absorptions per some period is Poisson distributed. The exponential character arises because the inter-successive-absorptions times are exponentially distributed, and the expected value between successive absorptions equals the expected time for the first absorption.

The conventional calculus-based proofs require some clarifications to address the issues raised by Bare. This is avoided in the proof of \S \ref{Sec: Deduccion 6} at the expense of making it long. This derivation maintains the approach used in the traditional proofs consisting of looking at losses caused by successive layers. It is meant to illustrate a viable procedure but not to be used in the classroom because simpler alternatives exist. It is noteworthy that this proof could be made much briefer by using product integrals\supercite{grossman1972non}, but it is not something students usually know about.

Among the derivations proposed in this paper, only that of \S \ref{Sec: Deduccion 7} does not refer explicitly to probability. However, it is implicitly implied through the inclusion of the reaction rate\supercite{GILLESPIE1992}. This derivation is based on the continuity equation which is used in many fields that are important to chemists and physicists.

\section{Conclusions}
In this work, seven derivations of the Lambert-Beer law were proposed. It is most likely that only one derivation will be shown in detail at the classroom, however, it is enriching to know different approaches  to achieve a more complete knowledge on the subject. 

\linespread{1.1} 
\printbibliography

\end{document}